\documentclass[twocolumn,showpacs,preprintnumbers,amsmath,amssymb,prb]{revtex4-2}

\usepackage{graphicx}
\usepackage{dcolumn}
\usepackage{bm}
\usepackage{amsmath}
\usepackage{amssymb}
\usepackage{ragged2e}
\usepackage{xcolor}
\usepackage[colorlinks=true,
linkcolor=blue, citecolor=blue, urlcolor=blue, filecolor=blue, anchorcolor=blue, bookmarks=true, bookmarksnumbered=true, pdftitle={Role of the dipolar tail in the dynamics of polar lattice gases}, pdfauthor={Gupta, Aramthottil, Santos}]{hyperref}

\usepackage{orcidlink}

\begin{document}


\title{Competing delocalisation and localisation induced by the dipolar tail in fragmented lattice gases}

\author{Sabhyata Gupta\,\orcidlink{0009-0003-3551-8651}}
\email{sabhyata.gupta@itp.uni-hannover.de}
\author{Adith Sai Aramthottil\,\orcidlink{0000-0002-4776-1523}}
\email{adith.sai@itp.uni-hannover.de}
\author{Luis Santos\,\orcidlink{0000-0001-7652-9574}}
\email{santos@itp.uni-hannover.de}
\affiliation{Institut f\"{u}r Theoretische Physik,
Leibniz Universit\"{a}t Hannover, 30167 Hannover, Germany}

\date{\today}

\begin{abstract}

We investigate how the long-range tail of dipole-dipole interactions modifies the dynamics of strongly interacting hard-core bosons in a one-dimensional lattice. In the nearest-neighbour model, large interactions generate an emergent Hilbert-space fragmentation into manifolds labelled by the number of nearest-neighbour bonds \(N_{\mathrm{NN}}\). The dipolar tail has a non-monotonic effect on this fragmented dynamics as the interaction-to-tunnelling ratio \(V/t\) is increased. At intermediate \(V/t\), it enhances delocalisation relative to the nearest-neighbour model by reducing the energetic detuning between neighbouring \(N_{\mathrm{NN}}\) manifolds and sustaining their hybridisation by hopping. At larger \(V/t\), the same tail lifts degeneracies within each manifold, beginning with the splitting associated with the next-nearest-neighbour bond number \(N_{\mathrm{NNN}}\), and promotes intra-manifold localisation and Hilbert-space shattering. The competition between these two effects produces a crossover from a delocalising to a localising role of the dipolar tail, visible in the late-time spatial inhomogeneity. This crossover occurs for nearly all initial states considered and persists at different fillings. Transverse confinement strengthens the short-range part of the dipolar tail and shifts the crossover to lower \(V/t\), bringing it into the regime relevant to current experiments with magnetic atoms and polar molecules.
\end{abstract}

\maketitle

\section{Introduction}

The breakdown of quantum thermalisation in isolated many-body systems is a central problem in non-equilibrium quantum physics~\cite{Deutsch1991,Srednicki1994,Tasaki1998,Rigol2008,DAlessio2016}. Besides integrability and disorder-induced many-body localisation~\cite{Basko2006,Oganesyan2007,Nandkishore2015,Abanin2019}, non-ergodic dynamics can arise from Hilbert-space fragmentation (HSF), in which kinetic constraints split conventional symmetry sectors into dynamically disconnected Krylov subspaces~\cite{Sala2020,Khemani2020,Moudgalya2022,MoudgalyaMotrunich2022}. Fragmentation has been studied in dipole-conserving spin chains~\cite{Sala2020,Khemani2020}, strongly interacting and tilted Hubbard systems~\cite{DeTomasi2019,Scherg2021,Kohlert2023,Bastianello2022}, and models with long-range interactions~\cite{Francica2023}, and signatures of restricted thermalisation have now been observed experimentally in several quantum-simulation platforms~\cite{Wang2025,Adler2024,Honda2025,Zhao2025}.

A particularly simple realisation is provided by hard-core bosons in a one-dimensional lattice with nearest-neighbour hopping \(t\) and repulsive nearest-neighbour interaction \(V\). In the limit \(V/t\rightarrow\infty\), resonant hopping conserves the number of nearest-neighbour bonds,
\(N_{\mathrm{NN}}=\sum_j n_jn_{j+1}\), and the Hilbert space fragments into sectors of fixed \(N_{\mathrm{NN}}\)~\cite{DeTomasi2019,Li2021}. At large but finite \(V/t\), these sectors are only approximately decoupled. For dipolar particles, however, the interaction extends beyond nearest neighbours. Previous work showed that, at sufficiently large \(V/t\), the dipolar tail resolves additional interaction-energy scales, beginning with the number of next-nearest-neighbour bonds \(N_{\mathrm{NNN}}\), and progressively shatters the original NN fragments into smaller subsectors, leading to strongly suppressed transport and disorder-free quasi-localisation~\cite{Li2020,Li2021}. The short-distance form of the dipolar interaction can furthermore be substantially modified by transverse confinement~\cite{Korbmacher2023}.

Here we show that the effect of the dipolar tail is non-monotonic as \(V/t\) is increased. At intermediate \(V/t\), the tail enhances delocalisation relative to the truncated NN model, whereas at larger \(V/t\) it promotes localisation and ultimately Hilbert-space shattering. The origin is a competition between two effects of the same long-range interaction. The tail reduces the energetic detuning between neighbouring \(N_{\mathrm{NN}}\) manifolds and thereby sustains their hybridisation by hopping, while at the same time lifting degeneracies within each manifold according to longer-range interaction energies, beginning with \(N_{\mathrm{NNN}}\). The former mechanism dominates at intermediate coupling, lowering the late-time spatial inhomogeneity below the NN value; above a crossover \((V/t)^*\), intra-manifold localisation prevails and the inhomogeneity becomes larger. This crossover occurs for nearly all initial states considered and persists at different fillings. Thus, the dipolar tail does not simply reinforce fragmentation. At intermediate coupling it can partially counteract it before driving the system towards Hilbert-space shattering at larger \(V/t\). 

This intermediate-coupling regime is also the one most relevant experimentally. Dipolar lattice gases with magnetic atoms typically operate at \(V/t\) of order unity to a few~\cite{Baier2016,Su2023}, while polar molecules provide stronger and electrically tunable interactions together with site-resolved detection and, more recently, independently controllable tunnelling and dipolar couplings~\cite{Rosenberg2022,Christakis2023,Mortlock2026,Carroll2025}. In practice, itinerant dipolar systems are naturally explored at finite interaction-to-tunnelling ratios, typically \(V/t\lesssim10\), rather than in the asymptotic strong-coupling regime. The crossover between tail-enhanced delocalisation and tail-induced localisation therefore occurs precisely in the parameter range in which fragmentation physics can be tested experimentally.





\section{Model and methods}
\label{sec:model}

We consider hard-core bosons on an open one-dimensional lattice of \(L\) sites containing \(N\) particles, described by
\begin{equation}
H=-t\sum_{j=1}^{L-1}\left(b_j^\dagger b_{j+1}+\mathrm{h.c.}\right)
+\sum_i\sum_{r>0}V_r\,n_i n_{i+r},
\label{eq:hamiltonian}
\end{equation}
where \(b_j^{(\dagger)}\) are hard-core bosonic operators at site $j$, and \(n_j=b_j^\dagger b_j\). The hopping amplitude \(t\) sets the energy scale, and we set \(\hbar=1\), so that time is measured in units of \(t^{-1}\).

For the bare dipolar model we take
$V_r=\frac{V}{r^3}$, 
where \(V\equiv V_1\) is the nearest-neighbour (NN) interaction strength. We compare this with a truncated NN model,
$V_r=V\,\delta_{r,1}$, 
which retains the same NN interaction while removing the dipolar tail. We also consider the effective interaction of a quasi-one-dimensional dipolar lattice derived in Ref.~\cite{Korbmacher2023}. Finite transverse confinement modifies the interaction at short distances, so that the resulting \(V_r\) no longer follows a pure power law. In our calculations we use the full confinement-modified interaction of Ref.~\cite{Korbmacher2023}, normalised such that \(V_1=V\). For convenience, we characterise the relative strength of its leading beyond-NN contribution by an effective exponent \(\beta_{\rm eff}\), defined through $V_2=\frac{V}{2^{\beta_{\rm eff}}}$.
In particular, we consider a confinement for which \(\beta_{\rm eff}=1.8\). As discussed below, the resulting enhancement of the short-range part of the dipolar tail has important consequences for the dynamics.


The real-time dynamics of dipolar lattice gases can be probed experimentally through site-resolved density measurements~\cite{Su2023,Christakis2023}, making the spatial inhomogeneity a natural diagnostic of relaxation~\cite{Li2021,Korbmacher2023}. Starting from a Fock state \(|\psi_0\rangle\) at \(\tau=0\), we evolve the state under Eq.~\eqref{eq:hamiltonian} using Krylov-subspace matrix exponentiation~\cite{AlMohy2011}. We quantify the degree of density inhomogeneity through
\begin{equation}
\eta(\tau)=
\frac{1}{2\bar n(1-\bar n)L}
\sum_j
\left|
\langle n_j\rangle(\tau)-\bar n
\right|,
\label{eq:eta}
\end{equation}
where \(\bar n=N/L\) is the mean filling. With this normalisation, \(\eta=1\) for any Fock state and \(\eta=0\) for a spatially uniform density profile. To characterise the long-time dynamics, we use the time-averaged inhomogeneity
\begin{equation}
\bar\eta=
\frac{1}{\tau_2-\tau_1}
\int_{\tau_1}^{\tau_2}d\tau\,\eta(\tau),
\end{equation}
with \(\tau_1=900/t\) and \(\tau_2=1000/t\).

To relate the dynamics to the structure of the many-body spectrum, we obtain the many-body eigenstates \(|\psi_\mu\rangle\) by exact diagonalisation. For each eigenstate, we calculate its inverse participation ratio in the Fock basis,
\begin{equation}
\mathrm{IPR}_\mu=
\sum_f
\left|
\langle f|\psi_\mu\rangle
\right|^4,
\end{equation}
where \(|f\rangle\) denotes a many-body Fock state. The IPR quantifies the extent to which an eigenstate is concentrated on a small number of Fock configurations. Large values indicate stronger Fock-space localisation, whereas small values correspond to states spread over many configurations.

To analyse the fragmentation structure, we additionally diagonalise Hamiltonians projected onto manifolds of fixed NN-bond number,
$H_n=P_nHP_n$,
where \(P_n\) projects onto the subspace with $N_{\mathrm{NN}}=\sum_j n_jn_{j+1}=n$. 
This allows us to distinguish processes that hybridise different \(N_{\mathrm{NN}}\) manifolds from localisation that develops within an individual manifold.

Unless stated otherwise, we consider chains of length \(L=16\) with open boundary conditions and half filling, \(N=8\). For the dynamical analysis, we evolve all \(4410\) Fock states belonging to the \(N_{\mathrm{NN}}=4\) manifold under the full \(N\)-particle Hamiltonian. Open boundary conditions are particularly relevant for the present problem because they suppress a singlon-assisted transport mechanism present in periodic systems. In the NN model, an isolated particle (singlon) can resonantly pass through a cluster of arbitrary length, displacing the cluster through the associated recoil process~\cite{DeTomasi2019}. With periodic boundary conditions, the same singlon can repeatedly circulate around the ring and undergo successive resonant encounters with the cluster, thereby producing a net cluster displacement and enhancing transport. In an open chain, by contrast, this recycling mechanism is interrupted once the singlon reaches a boundary. Open boundaries therefore suppress repeated singlon-assisted cluster motion and are also closer to the geometry of finite experimental lattice systems.



\begin{figure}[t!]
\includegraphics[width=\columnwidth]{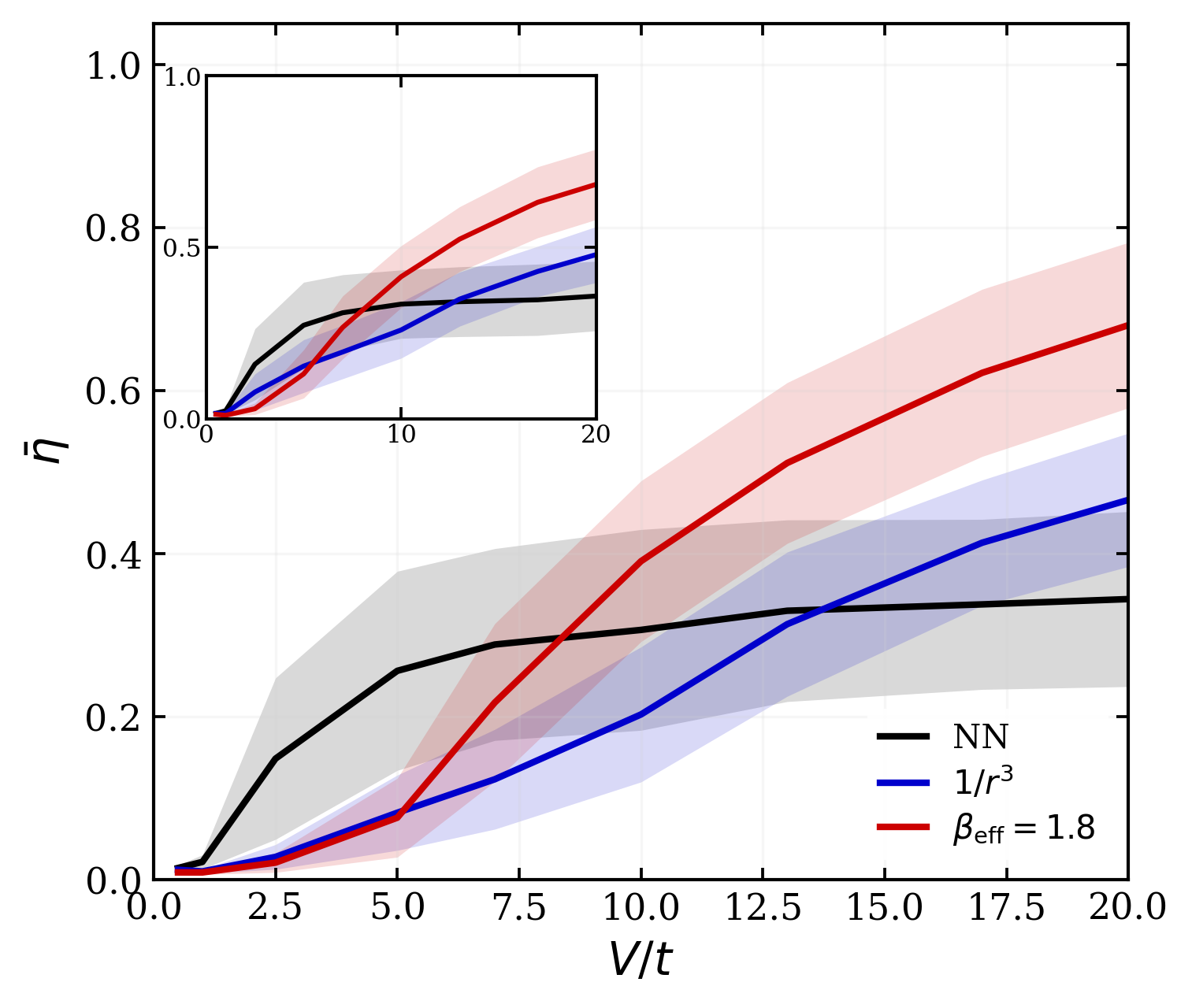}
\caption{Late-time inhomogeneity \(\bar{\eta}\) as a function of \(V/t\) for the
NN model (black), the bare dipolar interaction \(1/r^3\) (blue), and the modified
dipolar interaction with  \(\beta_{\mathrm{eff}}=1.8\) (red),
for an \(L=16\), \(N=8\) chain with open boundary conditions. Solid lines show
the average over all \(4410\) initial states in the \(N_{\mathrm{NN}}=4\)
manifold, while shaded regions indicate the corresponding standard deviation.
The NN result increases monotonically, whereas the dipolar models lie below it
at smaller \(V/t\) and above it at larger \(V/t\). The stronger NNN coupling for
\(\beta_{\mathrm{eff}}=1.8\) shifts the crossover to smaller \(V/t\) and produces
a larger inhomogeneity at strong coupling. The average is taken over
\(\tau\in[900,1000]\,\hbar/t\). The inset shows the same quantity averaged over
the earlier window \(\tau\in[190,200]\,\hbar/t\), demonstrating that the
crossover is already established at shorter, experimentally accessible times.}
\label{fig:crossover}
\end{figure}



\section{From tail-enhanced delocalisation to tail-induced localisation}
\label{sec:dynamics}

Figure~\ref{fig:crossover} shows the late-time inhomogeneity \(\bar{\eta}\), averaged over all \(4410\) initial Fock states in the \(N_{\mathrm{NN}}=4\) manifold, as a function of \(V/t\). We compare the NN model, the dipolar model with the bare \(1/r^3\) interaction, and the confinement-modified dipolar interaction characterised by \(\beta_{\mathrm{eff}}=1.8\). For the NN model, \(\bar{\eta}\) increases monotonically with \(V/t\), reflecting the progressive suppression of dynamics as the \(N_{\mathrm{NN}}\) manifolds become increasingly decoupled.

The dipolar models display a qualitatively different behaviour relative to the NN case. At smaller \(V/t\), the dipolar tail reduces the late-time inhomogeneity, so that \(\bar{\eta}\) lies below the NN result. As \(V/t\) increases, this trend reverses and the dipolar inhomogeneity eventually exceeds the NN value. For each initial state we define the crossover \((V/t)^*\) as the interaction strength at which
\[
\Delta\bar{\eta}
\equiv
\bar{\eta}_{\mathrm{dip}}-\bar{\eta}_{\mathrm{NN}}
\]
changes sign. The crossover is generic. It occurs for the overwhelming majority of the initial states considered~\footnote{\(4335\) out of the \(4410\) initial states exhibit the crossover for the \(1/r^3\) interaction, and \(4405\) out of \(4410\) for \(\beta_{\mathrm{eff}}=1.8\).}.

\begin{figure}[t]
\includegraphics[width=\columnwidth]{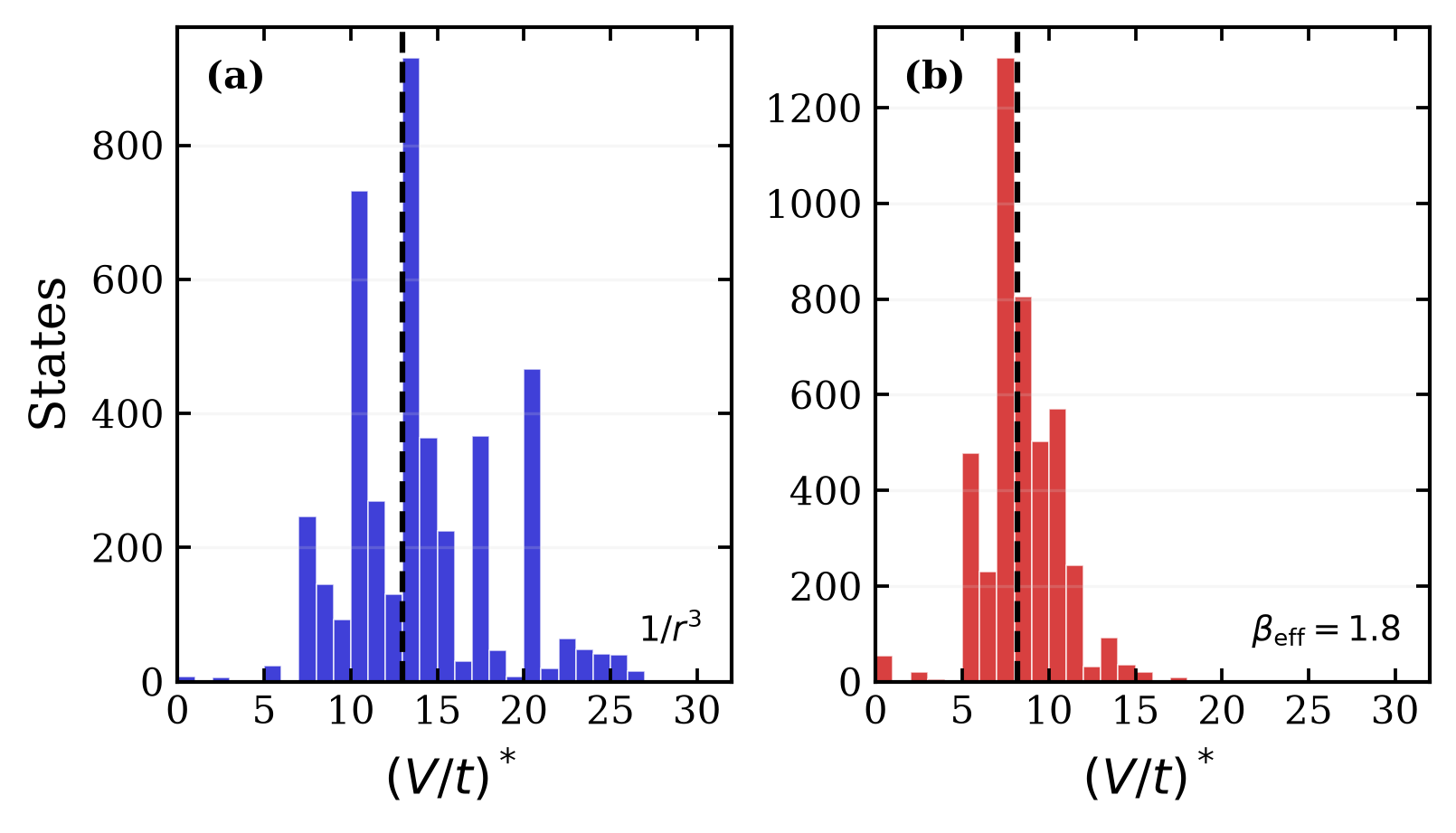}
\caption{Distribution of the crossover value \((V/t)^*\) over all initial states for
which \(\Delta\bar{\eta}\) changes sign, for (a) the \(1/r^3\) interaction and
(b) the modified dipolar interaction with \(\beta_{\mathrm{eff}}=1.8\). The
dashed vertical line in each panel marks the median, \((V/t)^*_{\mathrm{med}}=13.0\)
for \(1/r^3\) and \(8.1\) for \(\beta_{\mathrm{eff}}=1.8\). The larger NNN
coupling for \(\beta_{\mathrm{eff}}=1.8\) shifts the distribution towards smaller
\((V/t)^*\).}
\label{fig:vstar}
\end{figure}

The dependence of the crossover on the initial configuration is quantified in Fig.~\ref{fig:vstar}, which shows the distribution of \((V/t)^*\) over all initial states exhibiting a crossing. For the \(1/r^3\) interaction, the distribution is relatively broad, with a median value \((V/t)^*_{\rm med}\simeq13\), indicating a substantial dependence on the initial state. For the confinement-modified interaction with \(\beta_{\mathrm{eff}}=1.8\), the enhanced NNN coupling shifts the crossover to lower interaction strengths and considerably narrows the distribution, yielding a median \((V/t)^*_{\rm med}\simeq8.1\). The same ordering of the three models is already established at substantially shorter evolution times, as shown in the inset of Fig.~\ref{fig:crossover}.

We have verified that a similar crossover occurs at $L=12$ and other fillings as well, in particular at \(\bar n=1/3\) and \(\bar n=1/4\), and is therefore not specific to half filling. These results are not shown, and in the following we focus on \(\bar n=1/2\). We next turn to the microscopic origin of this non-monotonic role of the dipolar tail.



\begin{figure*}[t!]
\centering
\includegraphics[width=\textwidth]{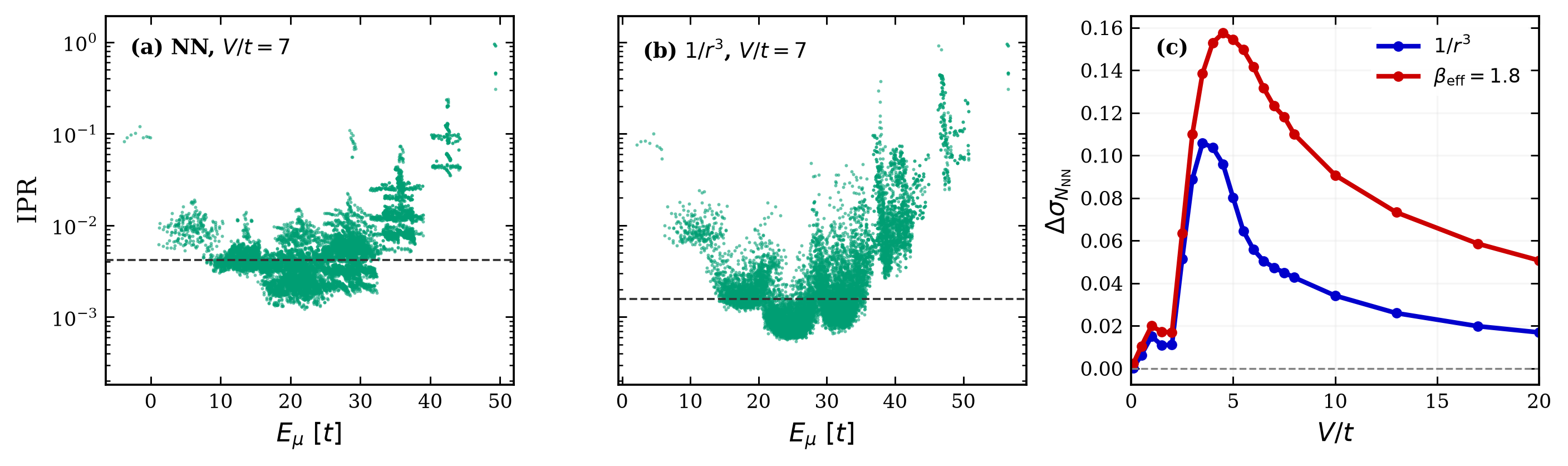}
\caption{Eigenstates and inter-manifold hybridisation for the \(L=16\), \(N=8\) chain. 
Panels (a) and (b) show the inverse participation ratio \(\mathrm{IPR}_\mu\) of the full-Hamiltonian eigenstates as a function of the eigenenergy \(E_\mu\) at \(V/t=7\), for (a) the NN model and (b) the bare \(1/r^3\) interaction. The dashed line indicates the median IPR in each panel. At this intermediate coupling, the dipolar interaction yields a smaller median IPR, showing that the eigenstates are spread over a larger number of Fock configurations than in the NN model. 
Panel (c) quantifies the associated inter-manifold hybridisation through the excess standard deviation of the NN-bond number, $\Delta\sigma_{N_{\rm NN}}$~(see text). Results are shown for the bare \(1/r^3\) interaction and for the confinement-modified interaction with \(\beta_{\mathrm{eff}}=1.8\), averaged over all initial states and over the time window \(\tau\in[900,1000]/t\).}

\label{fig:full_eigenstates}
\end{figure*}



\section{Mechanism}
\label{sec:mechanism}

The non-monotonic role of the dipolar tail identified in Sec.~\ref{sec:dynamics} originates from the competition between two distinct mechanisms. First, the tail reduces the energetic detuning between neighbouring \(N_{\mathrm{NN}}\) manifolds and thereby enhances their hybridisation by hopping, favouring delocalisation. Second, it lifts degeneracies within each \(N_{\mathrm{NN}}\) manifold by resolving differences in longer-range interaction energies, beginning with the \(N_{\mathrm{NNN}}\) contribution, and therefore promotes intra-manifold localisation. Both effects are present throughout the crossover, but their relative importance changes with \(V/t\). Inter-manifold hybridisation dominates at intermediate coupling, whereas intra-manifold localisation takes over at larger \(V/t\). We discuss these two mechanisms separately below.


\subsection{Inter-manifold hybridisation}

For the NN model, the interaction energy of a Fock configuration is simply \(V N_{\mathrm{NN}}\). Consequently, manifolds with different \(N_{\mathrm{NN}}\) separate linearly in energy as \(V\) increases. As long as neighbouring manifolds overlap energetically, NN hopping efficiently hybridises them, and the eigenstates of the full Hamiltonian carry appreciable weight over several \(N_{\mathrm{NN}}\) manifolds. Once the corresponding interaction-energy bands become sufficiently separated, this inter-manifold hybridisation is strongly suppressed and the dynamics becomes increasingly confined to the manifold containing the initial state.

The dipolar tail modifies this energetic separation. Its leading correction to the NN interaction energy is \(V_2N_{\mathrm{NNN}}\). Configurations belonging to neighbouring \(N_{\mathrm{NN}}\) manifolds can differ in \(N_{\mathrm{NNN}}\) in such a way that this contribution compensates part of their NN interaction-energy difference. The effective detuning between portions of adjacent manifolds is therefore reduced, maintaining inter-manifold resonances up to larger \(V/t\). Figure~\ref{fig:full_eigenstates} illustrates this directly through the many-body eigenstates \(|\psi_\mu\rangle\) of \(H\). At intermediate \(V/t\), the eigenstates of the dipolar model remain hybridised across several \(N_{\mathrm{NN}}\) manifolds, whereas in the NN model they have already reorganised into well-separated groups with nearly integer \(\langle N_{\mathrm{NN}}\rangle_\mu\).

To quantify this enhanced inter-manifold hybridisation, we evaluate the standard deviation
\begin{equation}
\sigma_{N_{\mathrm{NN}}}
=
\left(
\langle N_{\mathrm{NN}}^2\rangle
-
\langle N_{\mathrm{NN}}\rangle^2
\right)^{1/2},
\end{equation}
which measures the spread of the evolving state over different \(N_{\mathrm{NN}}\) manifolds. We then define the excess spread induced by the dipolar tail as
\begin{equation}
\Delta\sigma_{N_{\mathrm{NN}}}
=
\sigma_{N_{\mathrm{NN}}}^{\rm dip}
-
\sigma_{N_{\mathrm{NN}}}^{\rm NN}.
\end{equation}
Figure~\ref{fig:full_eigenstates}(c) shows \(\Delta\sigma_{N_{\mathrm{NN}}}\) for both dipolar interactions. It is positive and maximal at intermediate \(V/t\sim5\), demonstrating that the dipolar tail keeps the state distributed over a broader range of \(N_{\mathrm{NN}}\) manifolds than in the NN model. At larger \(V/t\), \(\Delta\sigma_{N_{\mathrm{NN}}}\) decreases towards zero as inter-manifold hybridisation becomes suppressed in both models. The confinement-modified dipolar interaction with \(\beta_{\mathrm{eff}}=1.8\) produces a larger \(\Delta\sigma_{N_{\mathrm{NN}}}\), showing that its stronger NNN coupling enhances the delocalising inter-manifold mechanism.




\begin{figure*}[t!]
\centering
\includegraphics[width=\textwidth]{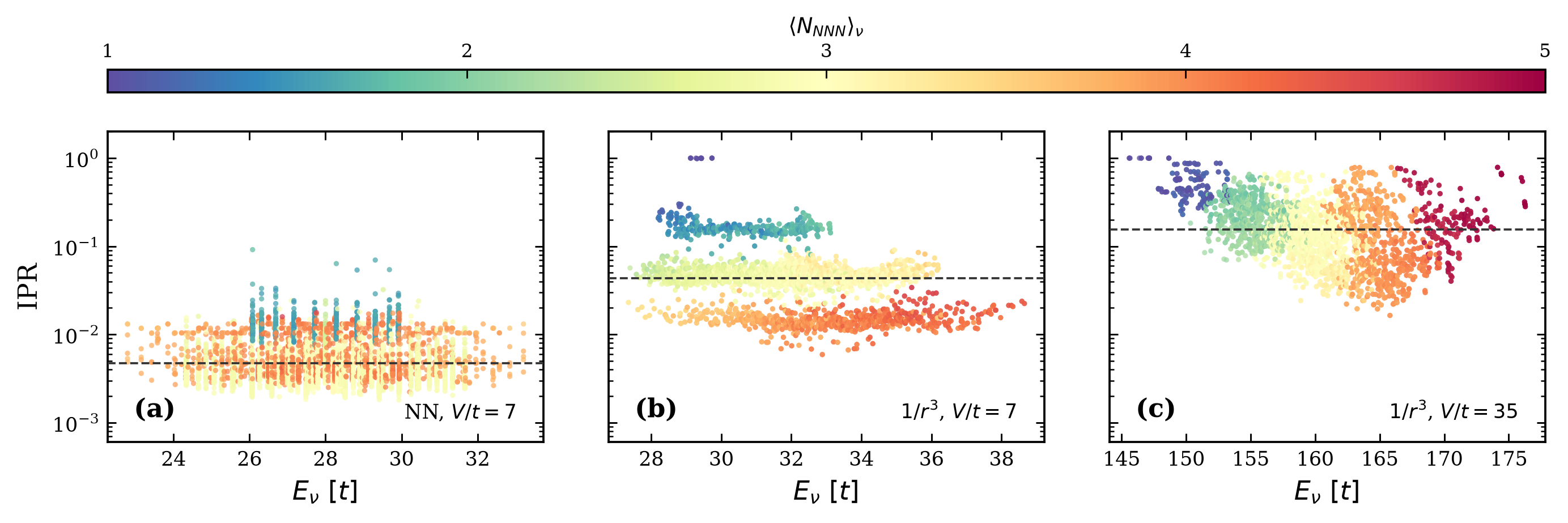}
\caption{Eigenstates of the Hamiltonian projected onto the \(N_{\mathrm{NN}}=4\) manifold for the \(L=16\), \(N=8\) chain, coloured by the mean NNN-bond number \(\langle N_{\mathrm{NNN}}\rangle_\nu=\sum_f|\langle f|\psi_\nu\rangle|^2 N_{\mathrm{NNN}}(f)\) (colour bar on top). Panels (a)--(c) show the inverse participation ratio as a function of the projected eigenenergy \(E_\nu\), for (a) the NN model at \(V/t=7\), (b) the \(1/r^3\) interaction at \(V/t=7\), and (c) the \(1/r^3\) interaction at \(V/t=35\), with the dashed line marking the median IPR in each panel. In the NN model all configurations within the manifold are degenerate, so the eigenstates remain delocalised in the Fock basis with low IPR independently of \(V/t\). The dipolar tail lifts this degeneracy according to the configurations' \(N_{\mathrm{NNN}}\) content and drives intra-manifold localisation, so the median IPR increases by roughly an order of magnitude from the NN model to \(1/r^3\) already at \(V/t=7\), and by a further factor of a few at \(V/t=35\). At this strong coupling the spectrum has reorganised into distinct \(N_{\mathrm{NNN}}\) sub-bands, visible as the vertical columns in panel (c), signalling Hilbert-space shattering.}
\label{fig:proj_eigenstates}
\end{figure*}


\subsection{Intra-manifold localisation}

Figure~\ref{fig:proj_eigenstates} shows the eigenstates of the Hamiltonian projected onto the \(N_{\mathrm{NN}}=4\) manifold, \(P_4HP_4\). In the NN model, all Fock configurations within this manifold have the same interaction energy \(4V\). The NN interaction therefore leaves their degeneracy intact; within the projected Hamiltonian it contributes only an overall energy shift. The corresponding eigenstates consequently remain strongly delocalised in the Fock basis, with low IPR, independently of \(V/t\), as shown in Fig.~\ref{fig:proj_eigenstates}(a).

The dipolar tail changes this situation qualitatively. Already its leading contribution, \(V_2N_{\mathrm{NNN}}\), lifts the degeneracy between Fock configurations with different \(N_{\mathrm{NNN}}\). As \(V/t\) increases, configurations with sufficiently different NNN interaction energies progressively decouple from one another, giving rise to eigenstates that become strongly localised in the Fock basis, with IPR values approaching unity. Such localised eigenstates already emerge at intermediate interaction strengths and become increasingly numerous as \(V_2\) grows relative to the kinetic bandwidth. This behaviour is visible in Fig.~\ref{fig:proj_eigenstates}(b), where eigenstates with different \(N_{\mathrm{NNN}}\) content coexist within the same energy range while displaying widely different IPR values. At still larger \(V/t\), the NNN interaction dominates the internal structure of the manifold and the spectrum reorganises into well-separated branches distinguished by \(N_{\mathrm{NNN}}\), as shown in Fig.~\ref{fig:proj_eigenstates}(c).

The competition between this intra-manifold localisation and the inter-manifold hybridisation discussed above determines the crossover scale \((V/t)^*\). The localisation mechanism becomes progressively stronger as \(V_2/t\) increases, because the NNN-induced energy splittings eventually overcome the kinetic processes that hybridise configurations within a given \(N_{\mathrm{NN}}\) manifold. This picture accounts qualitatively for two features of the numerical results. First, the crossover depends on which \(N_{\mathrm{NNN}}\) sectors are effectively sampled by a given initial state, contributing to the broad distribution of \((V/t)^*\) shown in Fig.~\ref{fig:vstar}. Second, increasing \(V_2\) shifts the onset of intra-manifold localisation to smaller \(V/t\). Consistently, the distribution of \((V/t)^*\) becomes narrower and its median decreases from approximately \(13\) for the bare \(1/r^3\) interaction to \(8.1\) for the confinement-modified interaction with \(\beta_{\mathrm{eff}}=1.8\).

\subsection{Initial period-four density wave}

The Fock state
\begin{equation}
|\psi_0\rangle=|1100110011001100\rangle
\label{eq:specialstate}
\end{equation}
provides a particularly transparent illustration of the tail-induced delocalisation mechanism. It belongs to the \(N_{\mathrm{NN}}=4\) manifold and has \(N_{\mathrm{NNN}}=0\). Moreover, every allowed single-particle hop takes the system from \(N_{\mathrm{NN}}=4\) to \(N_{\mathrm{NN}}=3\). Its relaxation therefore necessarily involves hybridisation between neighbouring \(N_{\mathrm{NN}}\) manifolds, making this configuration especially well suited for exposing the delocalising effect of the dipolar tail.

For the NN model, any directly connected configuration is separated from the initial state by an interaction-energy difference \(V\). The NNN interaction can substantially reduce this mismatch. For example,
\begin{equation}
|1{\bf 10}0110011001100\rangle
\longrightarrow
|1{\bf 01}0110011001100\rangle
\end{equation}
changes the bond numbers by
\begin{equation}
\Delta N_{\mathrm{NN}}=-1,
\qquad
\Delta N_{\mathrm{NNN}}=+2,
\end{equation}
so that, including the leading beyond-NN interaction, the corresponding energy detuning becomes
\begin{equation}
\Delta E = V-2V_2,
\end{equation}
instead of \(V\). For the bare dipolar interaction, \(V_2=V/8\), giving \(\Delta E=3V/4\), while transverse confinement can enhance \(V_2\) further and reduce the detuning even more.



\begin{figure}[t!]
\includegraphics[width=0.8\columnwidth]{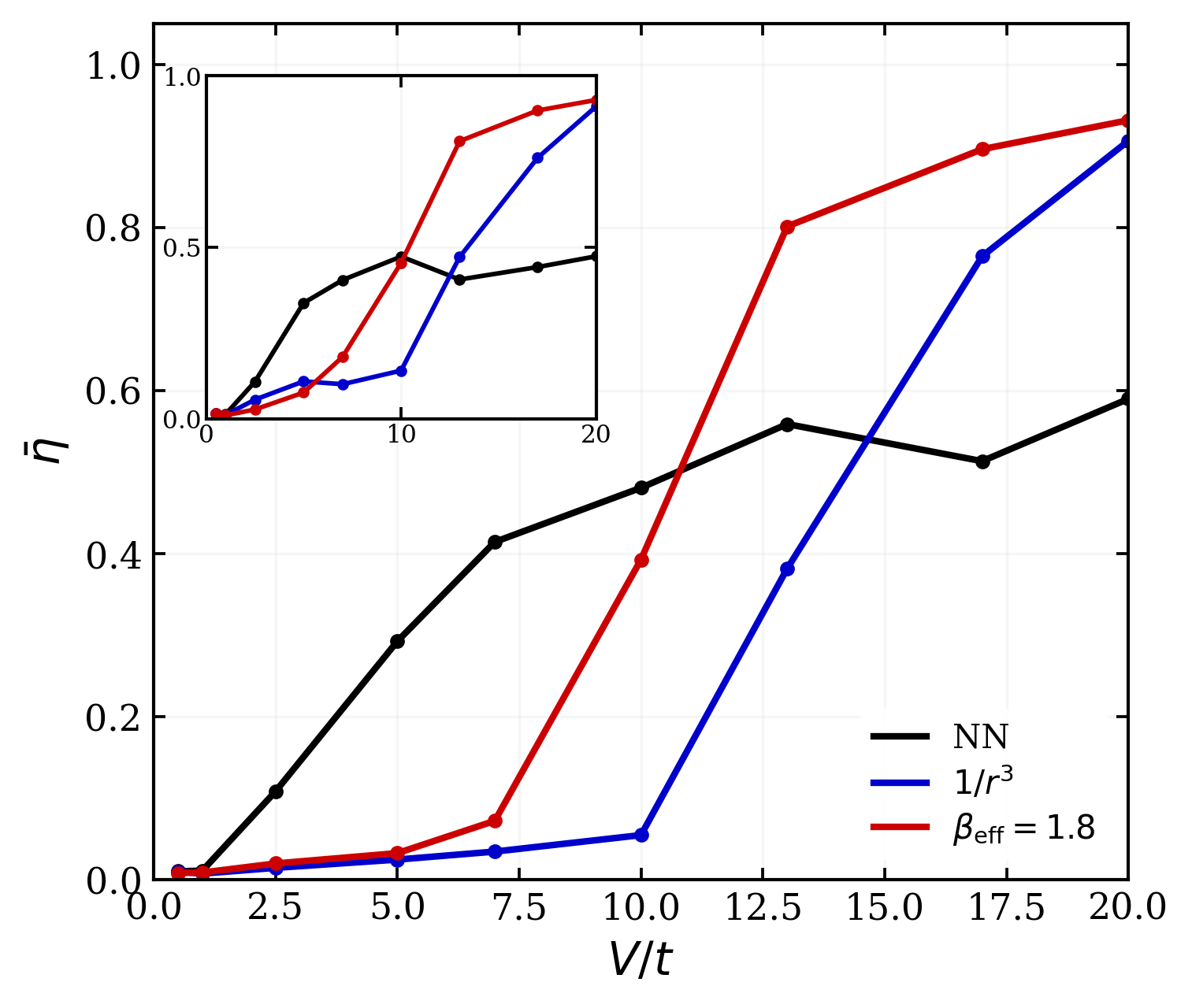}
\caption{Late-time inhomogeneity \(\bar{\eta}\) for the period-four initial state \(|1100\,1100\,1100\,1100\rangle\) in an \(L=16\), \(N=8\) chain, as a function of \(V/t\). Results are shown for the NN model (black), the bare dipolar interaction \(1/r^3\) (blue), and the modified dipolar interaction with effective exponent \(\beta_{\mathrm{eff}}=1.8\) (red). At intermediate \(V/t\), the dipolar tail reduces \(\bar{\eta}\) below the NN value, whereas at larger \(V/t\) it produces stronger localisation. Among the half-filled \(L=16\) states considered, this
configuration exhibits one of the largest reductions of \(\bar{\eta}\) relative to the NN model. The average is taken over \(\tau\in[900,1000]\,\hbar/t\). The inset shows the same quantity averaged over the earlier window \(\tau\in[190,200]\,\hbar/t\), confirming that the crossover is already established at shorter, experimentally accessible times.}
\label{fig:eta_1100}
\end{figure}


Importantly, the dipolar tail does not need to close the interaction-energy mismatch completely in order to enhance transport. The states in the neighbouring \(N_{\mathrm{NN}}=3\) manifold possess a finite kinetic bandwidth of order \(t\). By reducing the detuning from \(V\) to \(V-2V_2\), the dipolar tail brings a larger portion of this manifold into quasi-resonance with the initial configuration, thereby enhancing inter-manifold hybridisation relative to the NN model. This provides a simple microscopic interpretation of the reduced late-time inhomogeneity observed at intermediate \(V/t\).

At larger \(V/t\), the opposite effect of the tail becomes dominant. The interaction-energy splittings generated within the relevant \(N_{\mathrm{NN}}\) manifolds increasingly suppress hybridisation between Fock configurations, leading to stronger localisation than in the NN model. The competition between these two mechanisms produces the particularly clear crossover shown in Fig.~\ref{fig:eta_1100}.

This initial state is also experimentally attractive. Its simple period-four density pattern,
\(|1100\,1100\,\cdots\rangle\), can be prepared naturally using a period-four superlattice or an equivalent site-selective loading protocol. It therefore provides a convenient configuration for directly probing the crossover from tail-enhanced delocalisation at intermediate \(V/t\) to tail-induced localisation at larger \(V/t\).


\section{Conclusions}
\label{sec:conclusions}

The dipolar tail is commonly associated with the suppression of transport in lattice gases, since at strong interactions it resolves progressively finer interaction-energy scales, shatters the Hilbert-space fragments of the nearest-neighbour model, and ultimately localises the dynamics. Our results show that this picture is incomplete. At intermediate \(V/t\), before the shattering regime is reached, the dipolar tail instead enhances delocalisation relative to the truncated NN model and produces a more homogeneous late-time density profile. This non-monotonic behaviour originates from the competition between two effects generated by the same long-range interaction. On the one hand, the tail reduces the energetic detuning between neighbouring \(N_{\mathrm{NN}}\) manifolds and thereby sustains inter-manifold hybridisation. On the other hand, it lifts degeneracies within a fixed \(N_{\mathrm{NN}}\) manifold, beginning with the splitting associated with \(N_{\mathrm{NNN}}\), and thus promotes intra-manifold localisation. Both mechanisms coexist throughout the crossover, but their relative importance changes with \(V/t\). Their competition defines a crossover scale \((V/t)^*\), which we find to occur for nearly all initial states considered and to persist at different fillings.

The delocalising side of this competition is particularly relevant experimentally. Transverse confinement can substantially enhance the short-range part of the dipolar tail~\cite{Korbmacher2023}, thereby modifying the balance between inter- and intra-manifold processes. In our calculations, the median crossover shifts from \((V/t)^*\simeq13\) for the bare \(1/r^3\) interaction to \((V/t)^*\simeq8\) for the confinement-modified interaction characterised by \(\beta_{\mathrm{eff}}=1.8\). Dipolar lattice experiments typically operate at finite interaction-to-tunnelling ratios below or of order \(10\)~\cite{Su2023}, placing them naturally in a regime where tail-induced inter-manifold hybridisation can compete efficiently with fragmentation. Observing a late-time inhomogeneity below that of the corresponding NN model would therefore provide a direct signature of the dipolar interaction acting against, rather than reinforcing, localisation. The period-four state \(|1100\,1100\,\cdots\rangle\), which can be prepared using a period-four superlattice, provides a particularly transparent initial configuration for probing this effect.

More generally, our results identify the relative strength of interactions beyond nearest neighbours, and in particular the NNN scale \(V_2/t\), as a key parameter controlling the competition between inter-manifold hybridisation and intra-manifold localisation. The comparison between the bare and confinement-modified dipolar interactions suggests that analogous non-monotonic dynamics should arise in other power-law-interacting lattice systems whenever beyond-NN interactions are sufficiently strong to modify the energetic separation of neighbouring fragmentation manifolds.

\begin{acknowledgments}
We acknowledge the support of the Deutsche Forschungsgemeinschaft (DFG, German Research Foundation) -- Project-ID 274200144 -- SFB 1227 DQ-mat within the project A04, and under Germany's Excellence Strategy -- EXC-2123 Quantum-Frontiers -- 390837967.
\end{acknowledgments}

\bibliography{references}

\end{document}